\documentclass[a4paper,11pt]{article}

\usepackage{jinstpub} 

\usepackage{lineno}
\usepackage{hyperref}
\usepackage[figure]{hypcap}
\usepackage{placeins}

\usepackage{float}

\title{\boldmath Doping of a Borexino-like Liquid Scintillator with Tellurium-Diols}

\author[a,1]{Hans Th. J. Steiger,\note{Corresponding author.}}
\author[b]{Marco Beretta,}
\author[c,d]{Manuel Böhles,}
\author[e]{Alberto Garfagnini,}
\author[e]{Arsenii Gavrikov,}
\author[b]{Paolo Lombardi,}
\author[f]{Kai~Loo,}
\author[e]{Elena Pasini,}
\author[e]{Benedetta Rasera,}
\author[e]{Andrea Serafini,}
\author[c,d]{and Michael Wurm}

\affiliation[a]{Technical University of Munich, TUM School of Natural Sciences, Physics Department, \\ James-Franck-Str. 1, 85748 Garching, Germany}
\affiliation[b]{INFN, Sezione di Milano e Università degli Studi di Milano, Dipartimento di Fisica, Italy}
\affiliation[c]{Cluster of Excellence PRISMA$^+$ \\Staudingerweg 9, 55128 Mainz, Germany}
\affiliation[d]{Institute for Physics, Johannes Gutenberg University Mainz \\ Staudingerweg 7, 55128 Mainz, Germany}
\affiliation[e]{Dipartimento di Fisica e Astronomia dell’Universita di Padova and INFN Sezione di Padova, Padova, Italy}
\affiliation[f]{University of Jyväskylä, Department of Physics, Jyväskylä, Finland}
\emailAdd{hans.steiger@tum.de}

\abstract{
One promising approach for future neutrinoless double beta decay ($0\nu\beta\beta$) searches is the incorporation of candidate isotopes into liquid scintillator detectors. In this work, a sample of the high-performance 1,2,4-trimethylbenzene-based liquid scintillator used in the Borexino experiment was loaded with different concentrations of Te-diol compounds. To realize the loading, a modified water-free synthesis procedure in a non-acidic organic environment at room temperature was employed. The loaded scintillator mixtures were characterized with respect to their emission spectra, optical absorbance, light yield, and scintillation time profiles under $\alpha$ excitation. Within the experimental uncertainties, only comparatively small changes in the spectral emission shape and optical transmission were observed for Te-loadings up to 2\%. At the same time, a systematic reduction of the scintillation light yield with increasing Te concentration was measured. At 1\% Te-loading, an estimated light yield of approximately 8400\,photons/MeV$_{\mathrm{ee}}$ was obtained. Furthermore, the scintillation time profile measurements indicate systematically shorter effective decay time constants for increasing Te-loading, consistent with enhanced non-radiative de-excitation processes introduced by the Te-diol complexes. Overall, the results demonstrate that the investigated loading technique can be successfully applied to a pseudocumene-based liquid scintillator while preserving the principal scintillation characteristics of the system.
}


\arxivnumber{2405.05743} 

\begin{document}
\maketitle
\flushbottom

\section{Introduction}

The quest to detect neutrinoless double beta decay ($0\nu\beta\beta$) is a top priority in contemporary particle physics. Discovering this process, which violates lepton-number conservation, would not only reveal the absolute neutrino mass scale but also provide valuable insights into Grand Unified Theories and leptogenesis in the early universe~\cite{Zuber:2020kci}. 

To detect neutrinoless double beta decay, several experimental technologies have been developed over the last decades. The core idea is to incorporate into the detector’s active mass a ($0\nu\beta\beta$) candidate isotope: high-purity germanium detectors (LEGEND: $^{76}$Ge), loaded scintillators (KamLAND-Zen: $^{136}$Xe, SNO+: $^{130}$Te), and xenon-based time projection chambers in liquid or high-pressure gas form (EXO and NEXT: $^{136}$Xe) are among the current experiments. In particular, liquid scintillator (LS) detectors are highly sensitive instruments in the search for this ultra-rare process~\cite{SNOplus2021, Shimizu2019}. The detection method offers several advantages, including extremely low background radiation, flexibility in detector shape and size, and the ability to incorporate a large quantity of the $\beta\beta$-decaying isotope directly into the LS.

Significant effort has been devoted in recent years to the development of metal-loaded liquid scintillators, in particular those based on tellurium. The SNO+ collaboration has conducted extensive R\&D toward realizing a liquid scintillator loaded with compounds containing natural tellurium. The cocktail uses the widely employed organic solvent LAB (Linear alkylbenzene) and the fluor PPO (2,5-Diphenyloxazole)~\cite{Biller2017, Albanese2021, Anderson2021, Auty2023}, while the dopants are oligomers of Te-diols. These were prepared from an aqueous telluric acid solution with the addition of diols (typically 1,2-Butanediol) and stabilized with DDA (N,N-Dimethyldodecylamine)~\cite{Auty2023}. 

In addition to these developments, alternative synthesis routes have been explored. In particular, water-free or anhydrous approaches have been reported, for example by Ding et al.~\cite{Ding2023, Ding2025}, demonstrating that Te-diol compounds can also be prepared under non-aqueous conditions. Further studies, such as those by Suslov et al.~\cite{Suslov2022} and Liang et al.~\cite{Liang2025}, underline the broad interest in tellurium-loaded scintillator systems and highlight the diversity of chemical approaches under investigation.

While most current and future large-scale experiments rely on LAB or water-based liquid scintillators due to their favorable safety properties and scalability, pseudocumene (PC)-based scintillators remain a well-established benchmark system. The Borexino experiment has demonstrated outstanding radiopurity, high intrinsic light yield, and excellent timing performance in such a system~\cite{Giammarchi2014Borexino, Bellini2024BorexinoLegacy}. These properties make PC-based scintillators particularly suitable for fundamental studies of scintillation processes and for comparative investigations of loading techniques.

In this context, the present work is intended as a comparative and fundamental R\&D study. Rather than proposing a new detector concept, we investigate the behavior of tellurium loading in a high-performance pseudocumene-based scintillator derived from the Borexino experiment. This allows us to study the impact of Te-diol loading on key scintillator properties, including optical transparency, light yield, and timing response, in a system with well-characterized baseline performance.

The synthesis approach employed in this work is closely related to previously reported methods for Te-diol production~\cite{Auty2023, Ding2023}. However, it has been adapted to enable a fully water-free preparation and direct incorporation into a pseudocumene-based scintillator under non-acidic conditions. This provides a complementary perspective to existing LAB-based developments and extends the range of scintillator systems in which Te-loading has been successfully demonstrated.

\section{Te-diol Synthesis and Doping of 1,2,4-Trimethylbenzene-based Liquid Scintillator}

For the work described here, a sample of liquid scintillator from the Borexino experiment was provided by the Borexino Collaboration~\cite{Giammarchi2014Borexino}. The scintillator was handled under a protective nitrogen atmosphere and stored in light-tight glass containers prior to use. No modification of the original scintillator composition was performed before loading. The scintillator is based on the widely used solvent 1,2,4-trimethylbenzene (pseudocumene, PC; CAS: 95-63-6), combined with 1.5\,g/l 2,5-diphenyloxazole (PPO; CAS: 92-71-7) as fluor~\cite{Giammarchi2014Borexino}.

To realize the loading of the PC-based scintillator, the synthesis techniques described in~\cite{Auty2023} were adapted towards a water-free preparation in a non-acidic organic environment. This approach enables the formation of Te-diol complexes at room temperature without the need for external heating. Similar anhydrous synthesis strategies have been reported in the literature, for example by Ding et al.~\cite{Ding2023, Ding2025}. In the present work, commercially available chemicals of acceptable purity were used without further purification. The employed substances are summarized in \autoref{tab:chems}.

\begin{table}[h]
\centering
\resizebox{\textwidth}{!}{%
\begin{tabular}{|c|c|c|c|c|}
\hline 
Substance & Linear Formula & CAS Number & Purity & Supplier \\ 
\hline 
Orthotelluric acid (TeA) & H$_6$TeO$_6$ & 7803-68-1 & purum, $\geq$99.0\% & Sigma-Aldrich \\
\hline 
1,2-Butanediol (BD) & CH$_3$CH$_2$CH(OH)CH$_2$OH & 584-03-2 & purum, $\geq$98.0\% & Sigma-Aldrich \\ 
\hline 
N,N-Dimethyldodecylamine (DDA) & CH$_3$(CH$_2$)$_{11}$N(CH$_3$)$_2$ & 112-18-5 & $\geq$97.0\% & Sigma-Aldrich\\ 
\hline 
Benzene & C$_6$H$_6$ & 71-43-2 & for HPLC, $\geq$99.9\% & Sigma-Aldrich \\
\hline 
\end{tabular} 
}
\caption{Chemicals used for the synthesis of the Te-diol compounds. All substances were used as received without further purification.}
\label{tab:chems}
\end{table}

For the synthesis, orthotelluric acid was first ground to a fine powder and subsequently transferred into a round-bottom flask. The acid was neutralized with N,N-dimethyldodecylamine at room temperature under continuous stirring. Benzene was then added as an organic carrier medium. To form the Te-diol complexes, 1,2-butanediol was added dropwise using a burette (typical addition rate of order $\mathcal{O}$(0.1--1)\,ml/min). During the reaction, water is produced as a by-product and was continuously removed by bubbling dry nitrogen through the mixture. Over the course of several hours, the initially turbid suspension became fully transparent.

The molar ratios of the reactants were chosen to be 0.5:1:2 (DDA:TeA:BD), consistent with the stoichiometry reported in~\cite{Auty2023}. A slight excess of TeA was used to minimize the presence of unbound butanediol, which could negatively affect the optical transparency due to its limited solubility in PC. Residual solid components were removed by filtration using a Büchner funnel equipped with ash-free filter paper (retention: 2\,\textmu m). The final product was obtained by evaporating the benzene at room temperature under continuous stirring, resulting in a colorless, transparent, and viscous liquid.

It should be noted that benzene was chosen in this work primarily to facilitate efficient removal of the reaction water via azeotropic distillation under ambient conditions. This choice is motivated by the simplicity of the laboratory-scale procedure. However, due to the known toxicity of benzene, its use is not intended for large-scale applications. Alternative solvents such as toluene or cyclohexane are expected to provide comparable functionality and represent a more suitable option for scaled implementations. The present study should therefore be regarded as a proof-of-principle demonstration of the synthesis and loading technique.

To load the 1,2,4-trimethylbenzene-based scintillator, the synthesized Te-diol product was added in a quantity corresponding to a tellurium concentration of 2\% by mass in the final mixture. The corresponding mass fraction of the Te-diol compound is higher by a factor of approximately 2.67. The dopant dissolves completely in the scintillator within seconds. Lower concentrations were obtained by dilution of this stock solution with additional Borexino scintillator. All samples were extensively purged with dry nitrogen gas and stored under inert atmosphere in light-tight containers.

We note that the addition of Te-diol compounds slightly modifies the density of the scintillator and may therefore affect the effective PPO concentration. For the concentrations studied here, this effect is expected to be small and is not corrected for explicitly. Furthermore, no dedicated structural characterization (e.g.\ spectroscopic analysis) of the synthesized compound was performed; the chemical composition is assumed to be consistent with previously reported Te-diol systems based on the identical stoichiometry and synthesis pathway.

\FloatBarrier

\section{Characterization of Fundamental Scintillator Properties}
\label{sec:charLS}

Modern liquid scintillators for neutrino physics must fulfill increasingly demanding performance requirements. Particularly important are high optical transparency for scintillation light, high intrinsic light yield, emission spectra well matched to the spectral sensitivity of conventional photomultiplier tubes, and a well-characterized time profile of the scintillation emission. In addition, long-term chemical stability and low intrinsic radioactive contamination are essential for applications in rare-event searches.

The pseudocumene-based scintillator employed in the Borexino experiment has demonstrated exceptional performance in many of these aspects, including outstanding radiopurity, high light yield, and excellent timing characteristics~\cite{Giammarchi2014Borexino, Bellini2024BorexinoLegacy}. As such, it provides a well-established reference system for comparative studies of modified scintillator compositions.

In this section, the unloaded Borexino scintillator is compared with Te-loaded mixtures of different concentrations in order to evaluate the influence of Te-diol loading on fundamental scintillation properties. The investigated parameters include the spectral emission characteristics, optical absorbance, light yield, and scintillation time profile for $\alpha$-particle interactions.



\subsection{Light Emission Spectra}
\label{sec:emission}

To evaluate the effective emission spectra of the scintillator mixtures as observed by photosensors in a detector environment, samples with different Te-loadings were investigated using an Edinburgh~FS5 spectrofluorometer and compared to the unloaded Borexino scintillator. In order to account for self-absorption effects, all samples were measured pure and undiluted in a $10\times10\times40$\,mm$^3$ fused silica cuvette. The conventional geometry sample holder (Edinburgh SC-05) provides a comparatively long optical path through the scintillator. All samples were excited at (260\,$\pm$\,1)\,nm, close to the absorption maximum of pseudocumene at approximately 267\,nm. The excitation and emission slit widths were set to 1\,nm and 0.5\,nm, respectively. For comparison of the spectral shapes, all spectra were normalized to the maximum of the PPO emission peak.

\begin{figure}[htbp]
  \centering
  \includegraphics[width=1.0\textwidth]{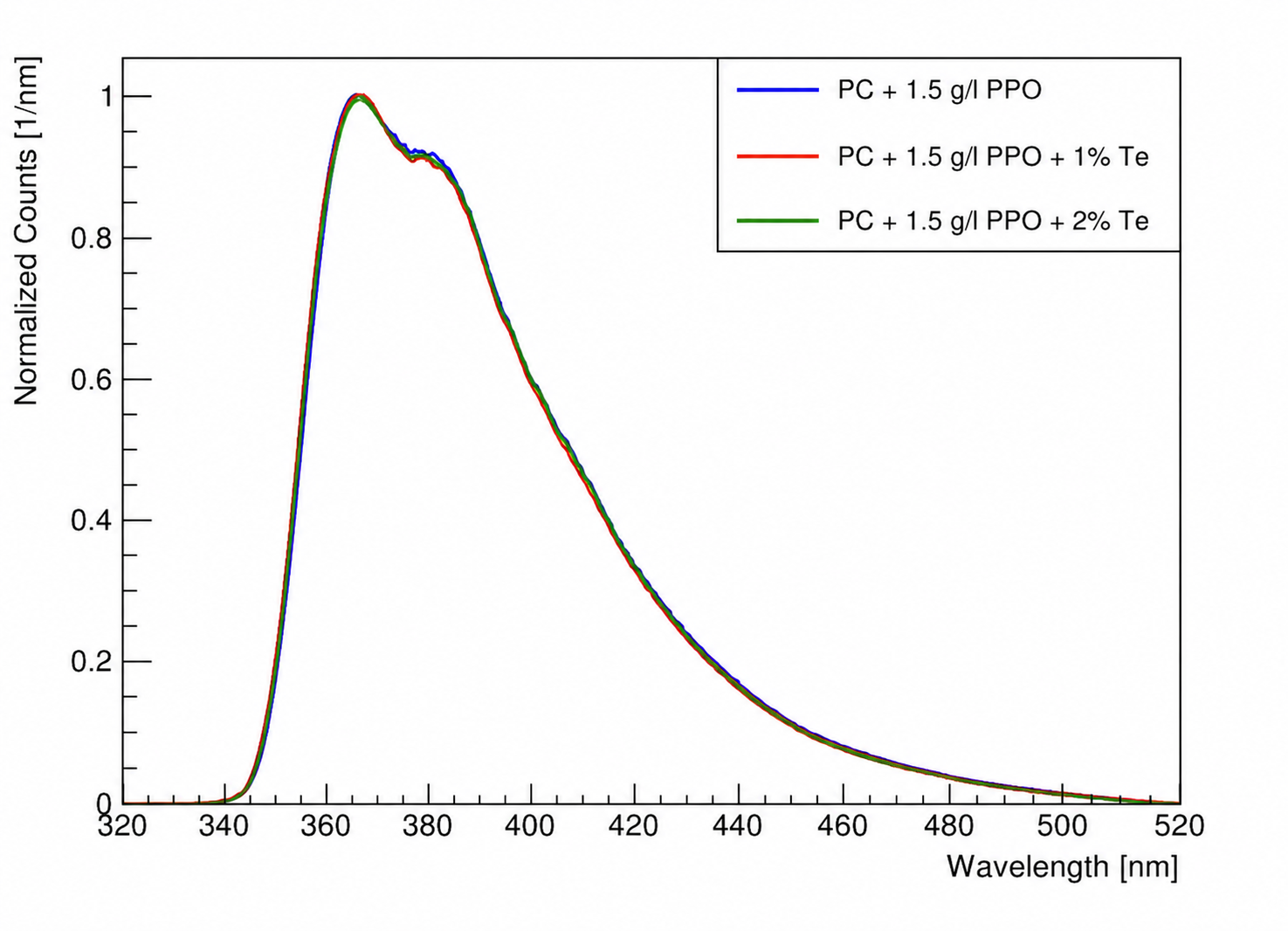} 
  \caption{Normalized effective emission spectra for the unloaded Borexino scintillator and samples loaded with 1\% and 2\% Te. All spectra are dominated by PPO emission. Within the measurement uncertainties, no substantial changes in spectral shape are observed.}
  \label{fig:EmissionSpectraCompare}
\end{figure}

The measured spectra are dominated by the characteristic PPO emission profile, exhibiting a pronounced maximum near 365\,nm and a broad tail extending towards larger wavelengths. Within the experimental uncertainties, only small differences between the unloaded and Te-loaded samples are observed. In particular, no substantial modification of the spectral shape is evident up to a Te-loading of 2\%. The effective emission spectra for the unloaded Borexino scintillator together with samples containing 1\% and 2\% Te-loading are shown in \autoref{fig:EmissionSpectraCompare}. While the spectral shape remains largely unchanged, the total amount of emitted scintillation light is affected by the Te-loading. This effect on the absolute light yield is discussed separately in \autoref{subsec:LightYield}.

\FloatBarrier

\subsection{Spectral Transmission of Light}
\label{sec:transparency}

\begin{figure}[htbp]
  \centering
  \includegraphics[width=1.0\textwidth]{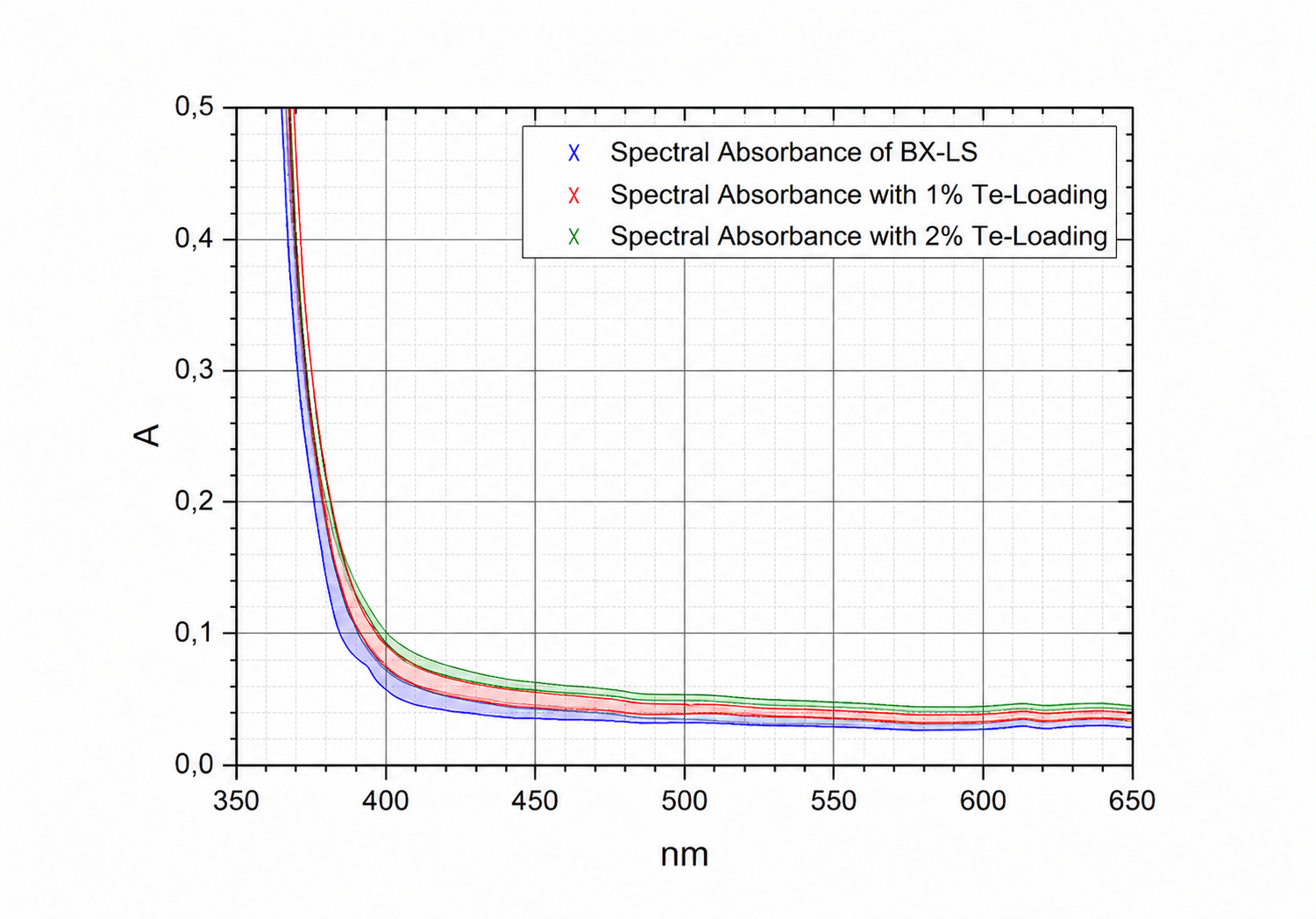} 
  \caption{Spectral absorbance of the unloaded Borexino scintillator compared with samples containing 1\% and 2\% Te-loading. The shaded bands represent the statistical uncertainty obtained from ten individual measurements per sample. Due to the finite cuvette length and instrumental limitations, additional systematic uncertainties are present but are not shown for clarity.}
  \label{fig:TransmissionSpectraCompare}
\end{figure}

To evaluate the optical transparency of the Te-loaded scintillator mixtures and to investigate possible degradation effects caused by increasing Te-loading, the spectral absorbance of the unloaded Borexino scintillator was compared with mixtures containing different concentrations of Te-diols. The measurements were performed using a Perkin Elmer Lambda 850+ UV/Vis spectrometer equipped with a rectangular fused-silica cuvette with an optical path length of 10\,cm. The resulting absorbance spectra are shown in \autoref{fig:TransmissionSpectraCompare}.

The measured spectra exhibit the characteristic strong increase in absorbance towards shorter wavelengths below approximately 400\,nm, while remaining comparatively flat in the visible wavelength region relevant for scintillation light detection. Within the experimental uncertainties, only small differences between the unloaded and Te-loaded samples are observed. In particular, the samples with Te-loadings between 0.125\% and 2\% do not show evidence for strong optical degradation or the appearance of additional pronounced absorption features in the wavelength range investigated.

A slight increase in absorbance for larger Te concentrations can be observed, especially at shorter wavelengths. However, these differences remain comparable to the statistical and systematic uncertainties of the measurement. The spectra for the unloaded scintillator together with samples containing 1\% and 2\% Te-loading are shown in \autoref{fig:TransmissionSpectraCompare}.

\FloatBarrier





\subsection{Light Yields}
\label{subsec:LightYield}

The light yield (LY), i.e.\ the number of scintillation photons emitted for a given energy deposition by a specific particle species, is one of the key parameters determining the performance of a liquid scintillator detector. In particular, the achievable energy resolution and vertex reconstruction precision strongly depend on the scintillation light output. 

In binary scintillator systems consisting of a solvent and a fluor, the light yield generally increases with increasing fluor concentration in the range of approximately 1\,g/l to 10\,g/l. In this regime, non-radiative losses in the solvent compete with the energy transfer from the solvent molecules to the fluor. For a given solvent-fluor combination, a characteristic critical fluor concentration exists, typically below 1\,g/l, at which approximately half of the maximum light yield is reached~\cite{BuckYeh, AberleChem}. Efficient solvent-to-fluor energy transfer therefore enables comparatively high light yields even at moderate fluor concentrations.

Absolute measurements of scintillation photon yields in liquid scintillators are experimentally challenging and subject to sizable systematic uncertainties~\cite{Bonhomme}. Consequently, light yields are commonly determined relative to a reference scintillator with known or well-characterized performance. In the present work, the unloaded Borexino scintillator was used as a relative reference sample.

\begin{figure}[H]
    \centering
    \includegraphics[width=0.53\textwidth]{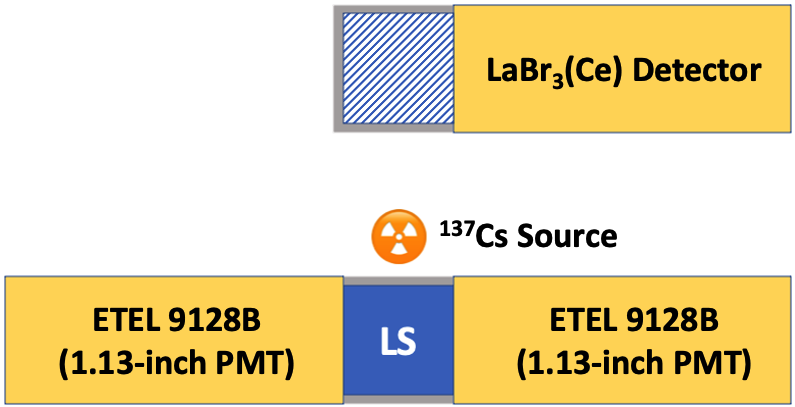}
    \caption{Schematic drawing of the light yield setup. A small highly reflective PTFE scintillator cell with UV-transparent glass windows is filled with the liquid scintillator sample and read out by two PMTs. The scintillator is irradiated by $\gamma$-rays from a $^{137}$Cs source. The coincidence with a LaBr$_3$(Ce) detector fixes the scattering geometry and thus the deposited energy in the liquid scintillator. Further details on the setup and analysis procedure are given in~\cite{SteigerSlow2024}.}
    \label{fig:LySetupMZ}
\end{figure}

The measurements were performed using the setup shown in \autoref{fig:LySetupMZ}, which is described in detail in~\cite{SteigerSlow2024}. A 10.9\,ml scintillator sample was filled under protective N$_2$ atmosphere into a highly reflective PTFE cell equipped with 2\,mm thin UV-transparent glass windows. Prior to and after filling, the samples were extensively purged with nitrogen gas to suppress oxygen-induced quenching effects. The scintillation light was detected using two 1.13\,inch PMTs (ETEL9128B) coupled directly to the cell windows.

During the measurements, the scintillator was irradiated using mono-energetic 662\,keV $\gamma$-rays from a $^{137}$Cs source (activity: $\sim$370\,kBq). To select a well-defined energy deposition in the scintillator, the setup was operated in coincidence with a 1.5\,inch $\times$ 1.5\,inch LaBr$_3$(Ce) detector. In the offline analysis, only events corresponding to the 185\,keV backscattering pseudo-peak in the crystal detector were selected, resulting in an energy deposition of approximately 477\,keV in the liquid scintillator~\cite{SteigerSlow2024, SteigerWbLS}.

\begin{figure}[htbp]
  \centering
  \includegraphics[width=0.8\textwidth]{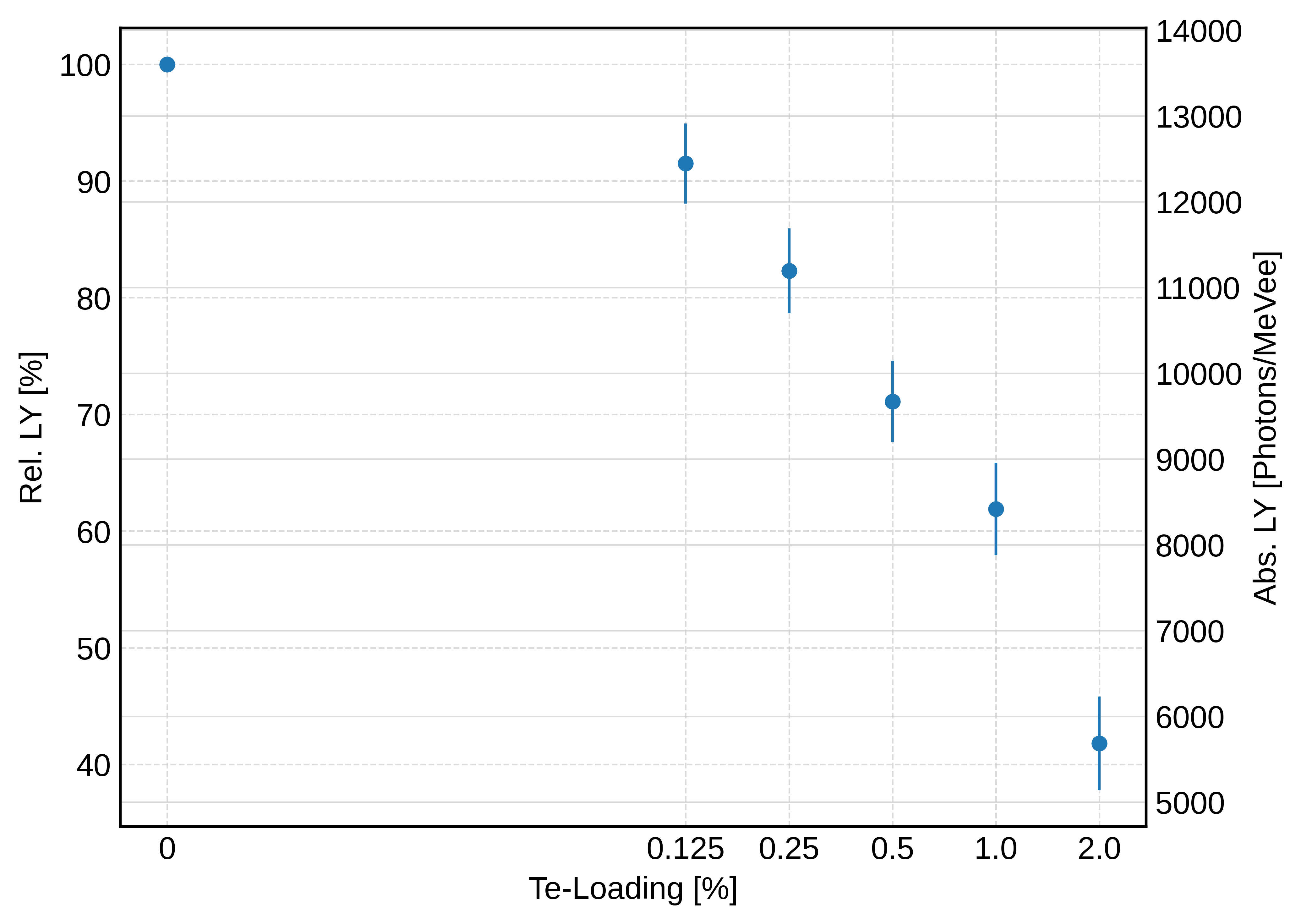} 
  \caption{Relative light yield as a function of Te-loading. Light yields exceeding $10^4$\,photons/MeV$_{\mathrm{ee}}$ are still achieved for moderate loading concentrations. For larger Te concentrations, increasing chemical quenching by the Te-diol complexes leads to a significant reduction of the scintillation light output in the PC/PPO scintillator system.}
  \label{fig:LyCompare}
\end{figure}

The measured relative light yields for different Te-loadings are summarized in \autoref{LyTab2} and shown graphically in \autoref{fig:LyCompare}. To estimate the combined statistical and systematic uncertainties, each sample was measured ten times on different days with freshly prepared and refilled cells. The observed spread of the measurements is attributed to effects such as PMT high-voltage stability, small laboratory temperature variations, residual oxygen contamination, and minor differences in sample preparation and detector cleaning.

A systematic decrease of the light yield with increasing Te concentration is observed. At a Te-loading of 1\%, the relative light yield decreases to approximately 62\% of the unloaded scintillator, corresponding to an estimated absolute light yield of roughly 8400\,photons/MeV$_{\mathrm{ee}}$. At 2\% loading, the light yield is reduced further to approximately 42\% of the unloaded sample. This reduction reflects the increasing contribution of non-radiative de-excitation channels introduced by the Te-diol complexes.

Assuming that the energy resolution is dominated by photon counting statistics, a first-order estimate can be obtained from $\sigma_E/E \propto 1/\sqrt{N}$, where $N$ is proportional to the number of detected scintillation photons. Based on the measured reduction from approximately 13600 \cite{Maneschg2014, Agostini2023CNO, Barbara} to 8400\,photons/MeV$_{\mathrm{ee}}$ at 1\% Te-loading, a degradation of the stochastic contribution to the energy resolution by approximately 25--30\% is expected. Note, that the estimated absolute value for the Borexino LY was obtained by scaling the measured relative light yields to the typical Borexino detector response of approximately 500\,photoelectrons/MeV reported in the literature~\cite{Maneschg2014, Agostini2023CNO}. The quoted values should therefore be regarded as approximate reference estimates \cite{Barbara}.

It should be emphasized that the experimentally measured quantity in this work is the relative light yield. The absolute values listed in \autoref{LyTab2} therefore represent estimated reference values obtained by scaling to typical Borexino scintillator performance parameters.

\begin{table}[h!]
\centering
\resizebox{\textwidth}{!}{%
\begin{tabular}{|c|c|c|c|c|c|c|}
\hline 
Te-Loading & 0\% & 0.125\% & 0.25\% & 0.5\% & 1.0\% & 2.0\%\\
\hline
Rel. LY [\%] & 100 & $91.5 \pm 3.42$ & $82.3 \pm 3.63$ & $71.1 \pm 3.50$ & $61.9 \pm 3.95$ & $41.8 \pm 4.01$\\ 
\hline
Estimated Abs. LY [Photons/MeV$_{\mathrm{ee}}$] & 13600 & $12444 \pm 466$ & $11193 \pm 494$ & $9670 \pm 476$ & $8418 \pm 538$ & $5685 \pm 546$\\ 
\hline
\end{tabular}%
}
\caption{Measured relative light yield values as a function of Te-loading. The estimated absolute values were obtained by scaling the measured relative light yields to typical Borexino scintillator performance parameters and are intended as approximate reference values.}
\label{LyTab2}
\end{table}

\FloatBarrier

\FloatBarrier

\FloatBarrier

\begin{figure}[htbp]
  \centering
  \includegraphics[width=0.95\textwidth]{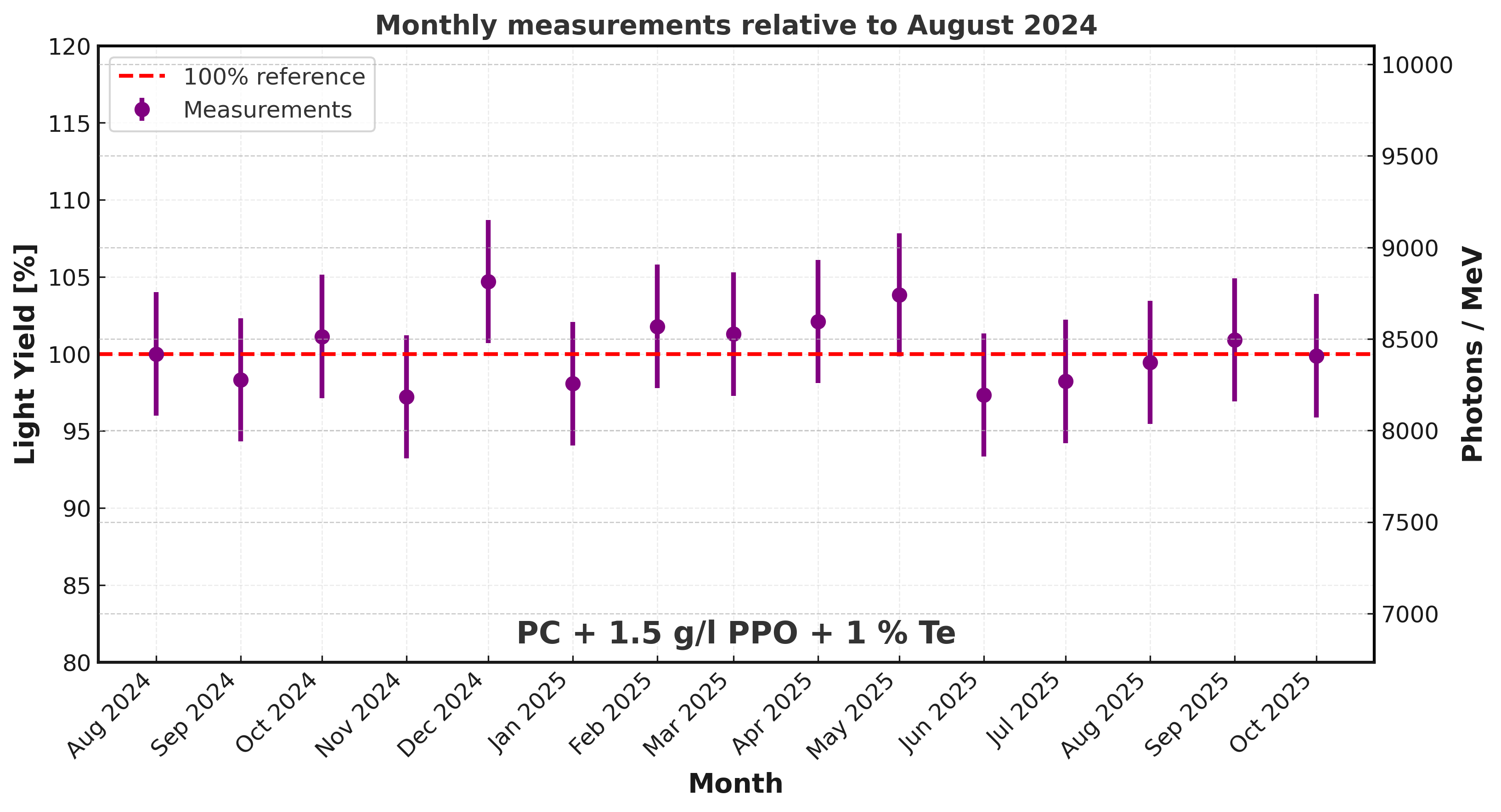}
  \caption{Repeated measurements of the relative light yield of a PC + 1.5\,g/l PPO + 1\% Te scintillator sample over a period exceeding one year. All values are normalized to the August~2024 measurement. Within the experimental uncertainties, no statistically significant long-term drift is observed.}
  \label{fig:LyStability}
\end{figure}

To investigate the long-term reproducibility and stability of the light yield setup, repeated measurements of a reference scintillator sample consisting of PC + 1.5\,g/l PPO + 1\% Te were performed over a period of more than one year. The measured light yield values, normalized to the initial August~2024 measurement, are shown in \autoref{fig:LyStability}. Within the experimental uncertainties, no significant long-term drift of the detector response or the scintillator performance is observed. The fluctuations remain at the level of only a few percent and are therefore substantially smaller than the light yield reduction induced by increasing Te-loading discussed above. These measurements demonstrate the robustness and reproducibility of the employed setup and support the interpretation that the observed reduction in light yield is dominated by the Te-loading itself rather than by instrumental instabilities or sample aging effects.

\FloatBarrier

\subsection{Light Emission Time Profile}
\label{sec:subsec}

The pulse shapes recorded by the photomultiplier tubes (PMTs) of a scintillation detector are determined not only by the detector electronics and photon sensor response, but fundamentally by the intrinsic fluorescence time profile of the scintillator itself. This time profile depends both on the scintillator composition and on the differential energy loss $dE/dx$ of the interacting particles. Detailed theoretical descriptions and experimental studies can be found in~\cite{Forster, Birks, Horrocks}. 

Since modifications of the scintillator composition may affect excitation transfer pathways and non-radiative de-excitation processes, it is important to investigate the influence of increasing Te-loading on the scintillation emission time profile. In the present work, the measurements focus specifically on $\alpha$-particle-induced scintillation signals and therefore provide information on the evolution of the intrinsic timing behavior of the scintillator system under increasing Te-loading. A dedicated investigation of pulse shape discrimination performance between different particle species is beyond the scope of this work.

\subsubsection{Light emission time profile setup}

\begin{figure}[H]
    \centering
    \includegraphics[width=0.45\linewidth]{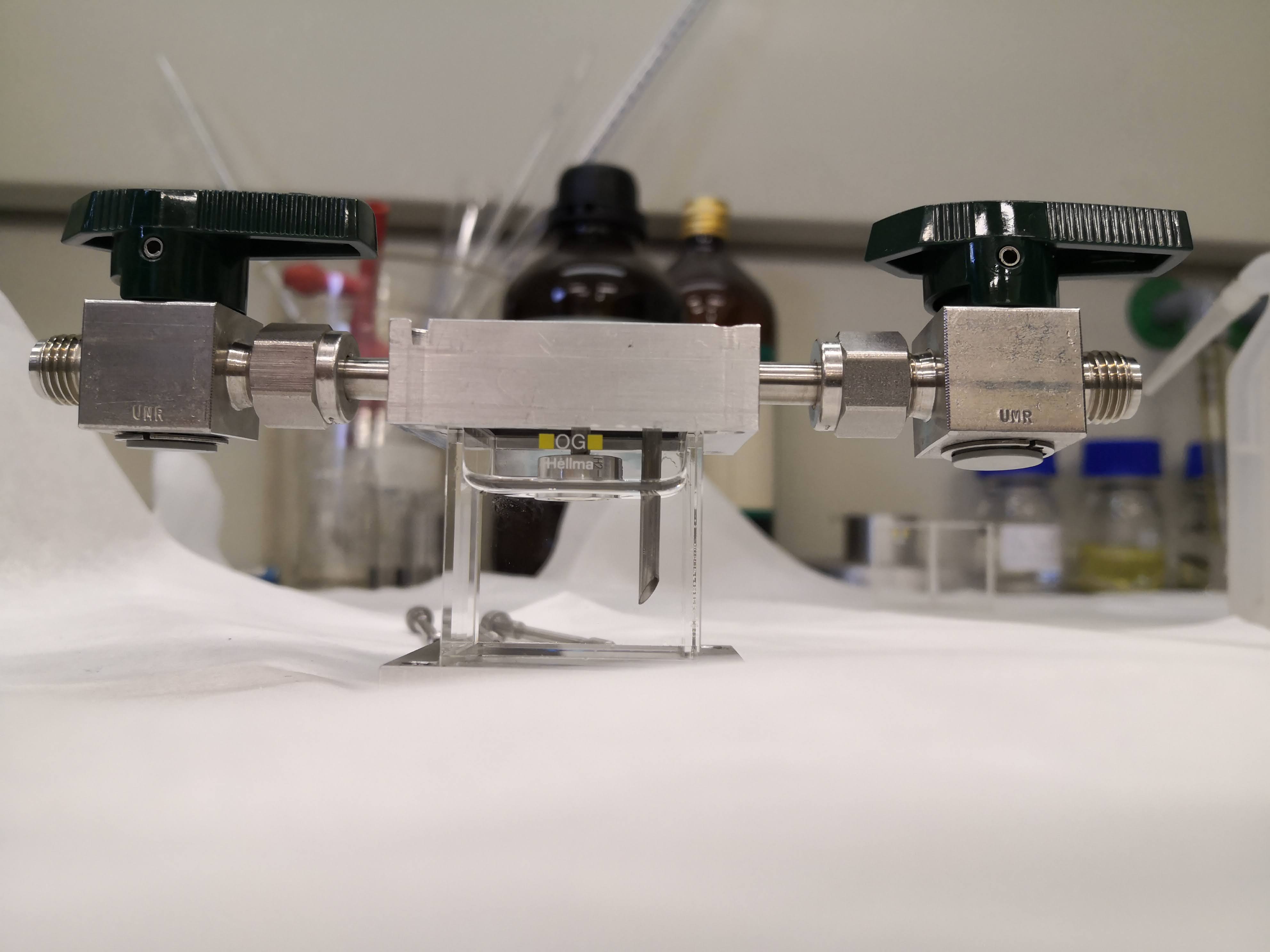}
    \includegraphics[width=0.45\linewidth]{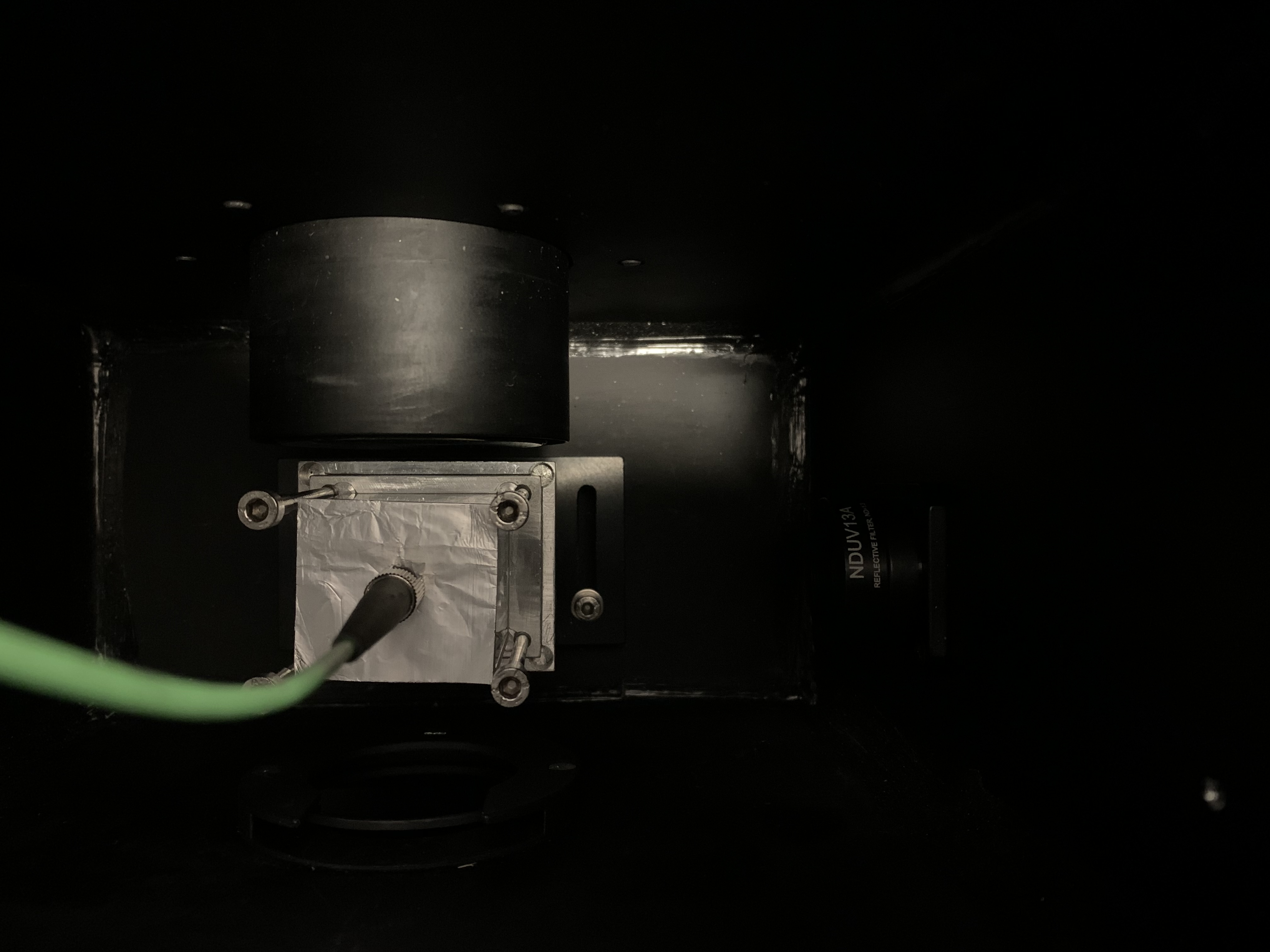}
    \caption{Left: UV-transparent scintillator cuvette equipped with a mounted $^{244}$Cm $\alpha$-source prior to nitrogen flushing. Right: Experimental arrangement used for the scintillation time profile measurements. Visible are the scintillator cuvette, the strongly coupled HL-PMT, the neutral density filter, and the weakly coupled LL-PMT operated in the single-photon regime.}
    \label{fig:Milan-setup}
\end{figure}

To determine the scintillation light emission time profiles, the experimental setup previously developed for the characterization of the JUNO liquid scintillator at the University of Milan was employed~\cite{Beretta2025}. The apparatus consists of a UV-transparent glass cuvette with dimensions of approximately $3\times3\times3$\,cm$^3$, filled with the scintillator sample under investigation. Radioactive sources can be mounted directly onto the vessel via a threaded source holder, as shown in the left panel of \autoref{fig:Milan-setup}.

To suppress oxygen-induced quenching effects, all samples were flushed with nitrogen gas for more than one hour after filling. During the measurements, a slight N$_2$ overpressure was maintained to ensure a stable inert atmosphere inside the cuvette.

The scintillation light was detected using two PMTs with different optical coupling strengths. The first PMT, referred to as the High-Level PMT (HL-PMT), was strongly coupled to the scintillator cell and provided the trigger signal. The second PMT, referred to as the Low-Level PMT (LL-PMT), was optically attenuated using a neutral density filter such that it operated in the single-photon detection regime. A detailed view of the detector arrangement is shown in the right panel of \autoref{fig:Milan-setup}.

This configuration enables the application of the well-established Time-Correlated Single Photon Counting (TCSPC) technique~\cite{Bollinger, Marrodan, HansPhD}. To verify the validity of the required single-photon condition, the charge spectrum of the weakly coupled PMT was continuously monitored during the measurements~\cite{Beretta2025}.

The instrumental response function (IRF) was characterized in situ using a pulsed EPL-405 diode laser with a typical pulse width of approximately 70\,ps. For the PMTs employed in this work, the laser signal can be approximated as a delta-like excitation pulse, thereby enabling a direct measurement of the detector response function~\cite{Beretta2025}. The intrinsic time resolution of the TCSPC system is approximately 0.45\,ns~\cite{Beretta2025}.

To suppress contamination from cosmic-ray-induced background events, the measurements were performed inside a muon veto system consisting of two EJ-200 plastic scintillator panels with dimensions of $500\,\mathrm{mm}\times500\,\mathrm{mm}\times20\,\mathrm{mm}$.

\subsubsection{Time profiles for $\alpha$-interactions with the liquid scintillator}

\begin{figure}[htbp]
  \centering
  \includegraphics[width=0.99\textwidth]{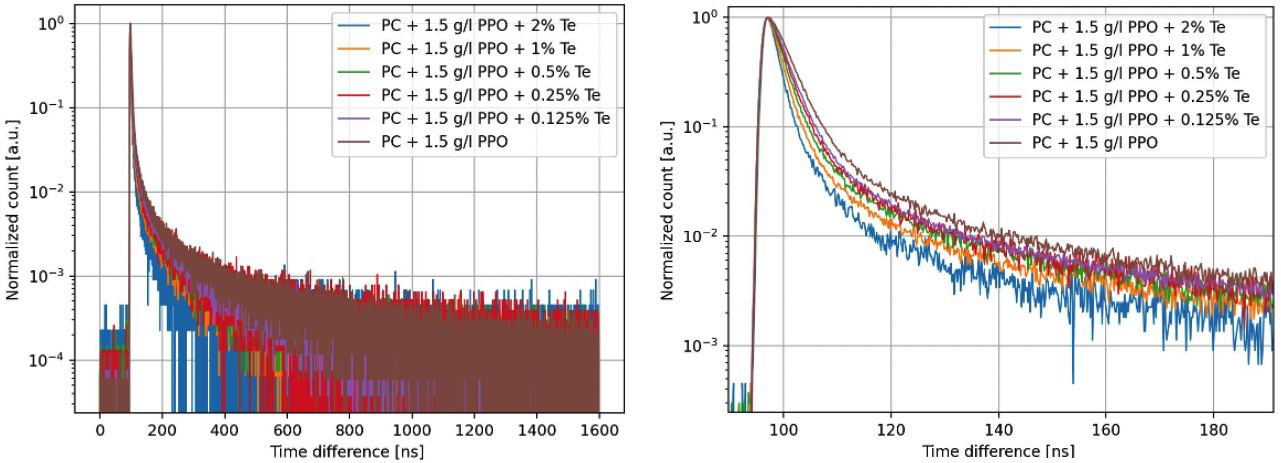} 
  \caption{Normalized scintillation time profiles for different Te-loadings under $\alpha$ excitation. The left panel shows the complete acquisition window, while the right panel focuses on the prompt decay region. Due to the reduced light yield at higher Te concentrations, the statistical precision decreases for strongly loaded samples. Nevertheless, the data quality remains sufficient to characterize the scintillation decay components in the first several hundred nanoseconds.}
  \label{fig:TimeProfile}
\end{figure}

For the time profile studies presented in this work, a $^{244}$Cm $\alpha$-source was deployed directly inside the scintillator vessel. After extensive nitrogen flushing, data acquisition was performed for scintillator mixtures with different Te-loadings. The measured scintillation time profiles are shown in \autoref{fig:TimeProfile} on a logarithmic scale. The left panel displays the full acquisition window of 1600\,ns, while the right panel provides a zoom into the first $\sim$100\,ns of the scintillation decay.

\FloatBarrier

To quantify the evolution of the scintillation decay characteristics, all datasets were fitted using a model consisting of four exponential decay components convoluted with the instrumental response function (IRF). The fit model is given in \autoref{Eq: FitModel}. A more detailed description of the setup, fitting procedure, and model assumptions is provided in~\cite{Beretta2025}.

\begin{equation}
	F_{\text{fit}}(t) = IRF(t - t_0) \otimes N \sum_{i=1}^4\frac{A_i}{\tau_i-\tau_r}\left(e^{-\frac{t-t_0}{\tau_i}}- e^{-\frac{t-t_0}{\tau_r}} \right)
	\label{Eq: FitModel}
\end{equation}

The resulting fit parameters are summarized in \autoref{PSDTab2}. With increasing Te-loading, a systematic reduction of all decay time constants is observed, while the relative amplitudes $A_i$ remain comparatively stable within uncertainties. This behavior is consistent with the increasing contribution of non-radiative de-excitation channels introduced by the Te-diol complexes, which modify the effective scintillation relaxation dynamics.

It should be emphasized that the present measurements were performed exclusively using $\alpha$-particle excitation. Consequently, the extracted decay constants characterize the timing behavior for highly ionizing particles and do not by themselves allow a direct quantitative assessment of pulse shape discrimination performance between $\alpha$ and $\beta/\gamma$ interactions.

\begin{table}[h!]
\centering
\resizebox{\textwidth}{!}{%
\begin{tabular}{|c|l|l|l|l|l|l|}
\hline 
Te-Loading & 0\% & 0.125\% & 0.25\% & 0.5\% & 1.0\% & 2.0\%\\
\hline
$A_1$\,[\%] & $68.15 \pm 0.15$ & $69.17 \pm 0.16$ & $66.75 \pm 0.31$ & $67.47 \pm 0.19$ & $68.10 \pm 0.43$ & $66.86 \pm 0.43$\\ 
$A_2$\,[\%] & $14.19 \pm 0.18$ & $14.14 \pm 0.14$ & $14.68 \pm 0.27$ & $14.44 \pm 0.21$ & $14.68 \pm 0.30$ & $17.30 \pm 0.30$\\   
$A_3$\,[\%] & $9.60 \pm 0.16$ & $9.76 \pm 0.11$ & $10.12 \pm 0.23$ & $9.76 \pm 0.17$ & $9.27 \pm 0.26$ & $8.79 \pm 0.26$\\ 
$A_4$\,[\%] & $8.20 \pm 0.29$ & $7.20 \pm 0.24$ & $8.80 \pm 0.48$ & $8.60 \pm 0.33$ & $8.00 \pm 0.59$ & $7.20 \pm 0.59$ \\ 
\hline
$\tau_1$\,[ns] & $3.212 \pm 0.001$ & $2.789 \pm 0.004$ & $2.549 \pm 0.008$ & $2.341 \pm 0.001$ & $2.010 \pm 0.001$ & $1.501 \pm 0.001$\\ 
$\tau_2$\,[ns] & $15.05 \pm 0.29$ & $12.59 \pm 0.20$ & $11.38 \pm 0.35$ & $11.48 \pm 0.25$ & $9.62 \pm 0.25$ & $6.59 \pm 0.24$\\
$\tau_3$\,[ns] & $74.10 \pm 2.70$ & $67.57 \pm 1.12$ & $65.36 \pm 3.07$ & $61.30 \pm 2.40$ & $55.90 \pm 3.20$ & $46.92 \pm 3.20$\\ 
$\tau_4$\,[ns] & $478.9 \pm 30.3$ & $470.0 \pm 30.0$ & $482.7 \pm 40.2$ & $448.9 \pm 30.4$ & $385.0 \pm 51.6$ & $399.8 \pm 51.6$\\
\hline
\end{tabular}%
}
\caption{Fit parameters obtained from the scintillation time profile measurements under $\alpha$ excitation for different Te-loadings. The quoted uncertainties include statistical and estimated systematic contributions. Increasing Te-loading leads to systematically shorter effective decay time constants.}
\label{PSDTab2}
\end{table}

\FloatBarrier

\section{Conclusions}
\label{sec:conclusions}

Metal-loaded liquid scintillators remain a promising technology for future searches for neutrinoless double beta decay. Experiments such as SNO+ and KamLAND-Zen have demonstrated the feasibility of isotope loading in large liquid scintillator detectors and motivated continued research on alternative loading techniques and scintillator systems.

In the present work, a Borexino-like pseudocumene-based liquid scintillator was successfully loaded with Te-diol compounds using a modified synthesis procedure derived from previously reported approaches. The applied method enables a fully water-free preparation in a non-acidic organic environment and allows the synthesis of the Te-diol complexes at room temperature without external heating.

A detailed characterization of the loaded scintillator mixtures was performed, including measurements of emission spectra, optical absorbance, light yield, and scintillation time profiles under $\alpha$ excitation. The measured emission spectra remain strongly dominated by PPO fluorescence for all investigated loading concentrations up to 2\% Te. Likewise, only comparatively small changes in the optical absorbance were observed within the experimental uncertainties, and no pronounced additional absorption structures appeared in the wavelength region relevant for scintillation light detection.

At the same time, a systematic reduction of the scintillation light yield with increasing Te-loading was observed. At 1\% Te concentration, the relative light yield decreases to approximately 62\% of the unloaded scintillator, corresponding to an estimated light yield of roughly 8400\,photons/MeV$_{\mathrm{ee}}$. While this reduction is significant, light yields exceeding $10^4$\,photons/MeV$_{\mathrm{ee}}$ are still achieved for moderate loading concentrations.

The scintillation time profile measurements show a systematic reduction of the effective decay time constants with increasing Te-loading, while the relative amplitudes of the decay components remain comparatively stable. This behavior is consistent with increasing contributions from non-radiative de-excitation pathways introduced by the Te-diol complexes. However, it should be emphasized that the presented timing studies were performed exclusively under $\alpha$ excitation and therefore do not by themselves provide a direct quantitative evaluation of pulse shape discrimination capabilities for different particle species.

Overall, the presented results demonstrate that the Te-loading technique investigated in this work can be successfully applied to a pseudocumene-based liquid scintillator while preserving the principal scintillation characteristics of the system. The study therefore provides a complementary perspective to existing LAB-based developments and contributes to the broader understanding of Te-loaded scintillator systems.


\acknowledgments

We gratefully acknowledge seed funding from the Munich Cluster of Excellence~ORIGINS (DFG, German Research Foundation under Germany’s Excellence Strategy~–~EXC-2094~–~390783311) and the support of the Detector Laboratory of the Mainz Cluster of Excellence PRISMA$^+$.\\ Special thanks go to the staff of the Detector Division, in particular Dr. Quirin Weitzel, for their practical assistance. We also warmly thank Dr. Mihail Mondeshki (Chemistry Department, JGU Mainz) and his team for their chemical expertise, constructive interdisciplinary collaboration, and numerous insightful discussions. We would like to thank the chemotechnical employee of the Mainz Institute for Physics, Joachim Strübig, and the glass apparatus manufacturer Rainer Jera.\\
We are grateful to the University of Milan and to INFN of Milan which host the fluorescence time profile setup in the Liquid Scintillator laboratory. We would also thank Dr. Barbara Caccianiga and Fatima Horuia for their help for this work.\\
For countless detailed and inspiring discussions we would like to thank especially Prof. Dr. Franz von Feilitzsch\mbox~(TUM), Dr. Gioacchino Ranucci\mbox~(INFN Milano), Prof. Dr. Caren Hagner (University of Hamburg and DESY),  Prof. Dr. Gabriel Orebi-Gann\mbox~(UC Berkeley and LBNL) and Dr. Minfang Yeh (BNL). Moreover, we would like to thank Prof. Dr. Peter Fierlinger\mbox~(TUM) and PD Dr. habil. Joachim Diener\mbox~(TUM) for the support of the work carried out in Munich.





\bibliographystyle{JHEP.bst}
\bibliography{main.bib}








\end{document}